\documentclass[runningheads]{llncs}
\usepackage{amsmath,amssymb,amsfonts}
\usepackage{algorithmic}
\usepackage{graphicx}
\usepackage{textcomp}
\usepackage{xcolor}
\usepackage{enumitem}
\usepackage{mathrsfs}

\usepackage{subfigure}
\usepackage{url}
\usepackage{pifont}

\usepackage{graphbox}

\usepackage{multirow}
\usepackage{textcase}

\usepackage{listings}
\usepackage{xcolor}
\definecolor{commentgreen}{RGB}{2,112,10}
\definecolor{eminence}{RGB}{108,48,130}
\definecolor{weborange}{RGB}{255,165,0}
\definecolor{frenchplum}{RGB}{129,20,83}

\newcommand{\fname}{CoGenT}
\newcommand{\mname}{pseudo-missing}
\begin{document}

\newcommand{\etal}{\textit{et al}.}
\newcommand{\etc}{\textit{etc}}
\newcommand{\ie}{\textit{i}.\textit{e}.}
\newcommand{\eg}{\textit{e}.\textit{g}.}

\title{CoGenT: A Content-oriented Generative-hit Framework for Content Delivery Networks\thanks{Supported by organization x.}}
%
%
\author{Peng Wang\inst{1}\orcidID{0000-1111-2222-3333} \and
Yu Liu\inst{1} \and
Ziqi Liu\inst{1} \and
Mingyang Wang\inst{1} \and Ke Liu\inst{1} \and Ke Zhou\inst{1} \and Zhihai Huang\inst{2}
}
\authorrunning{P. Wang et al.}
%
\institute{Huazhong University of Science and Technology, Wuhan, China \email{\{wp\_hust,liu\_yu,liuziqi,wangmy,liu\_ke,zhke\}@hust.edu.cn} \and
Tencent Technology (Shenzhen) Co., Ltd. Shenzhen, China
\email{tommyhuang@tencent.com}}
\maketitle              
\begin{abstract}
The service provided by content delivery networks (CDNs) may overlook content locality, leaving room for potential performance improvement. In this study, we explore the feasibility of leveraging generated data as a replacement for fetching data in missing scenarios based on content locality. Due to sufficient local computing resources and reliable generation efficiency, we propose a content-oriented generative-hit framework (\fname) for CDNs. \fname~utilizes idle computing resources on edge nodes to generate requested data based on similar or related cached data to achieve hits. Our implementation in a real-world system demonstrates that~\fname~reduces the average access latency by half. Additionally, experiments conducted on a simulator also confirm that \fname~can enhance existing caching algorithms, resulting in reduced latency and bandwidth usage.

\end{abstract}
\section{Introduction}
Content Delivery Networks (CDNs) are extensively employed across numerous edge nodes, utilized to store various types of data such as texts, images, videos, blocks,~\etc. Users can access data stored in CDNs to enjoy high-quality services with low latency. However, if the requested data is not available on CDNs, it needs to be fetched from a remote data center, resulting in increased latency. To mitigate this problem, classic caching algorithms~\cite{wang2021cost,song2023halp} are widely implemented in CDNs to maximize the caching of requested data based on data locality~\cite{song2020learning}. However, users may prioritize the content itself rather than the specific data cached in CDNs. For example, users are willing to change the video resolution from high to low when encountering buffering, which enables smoother video playback. This behavior may lead to a situation where CDNs cache videos with the same content but different resolutions, which can strain the CDNs' capacity to accommodate other data. To avoid this situation while improving content-oriented services by caching more comprehensive content, it is necessary to enhance the caching scheme of CDNs according to content locality.

\begin{figure*}[t]
    \centering
    \subfigure[City\#1]{
	\includegraphics[width=0.31\linewidth]{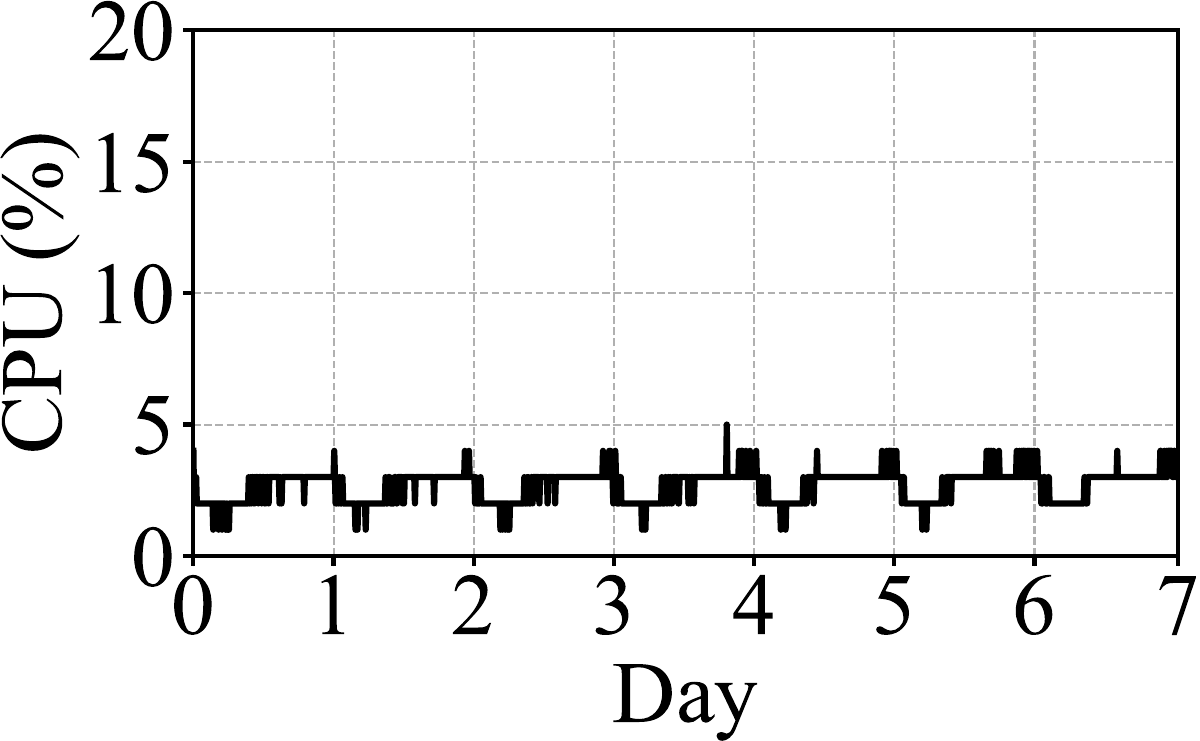}
    }
    \subfigure[City\#2]{
	\includegraphics[width=0.31\linewidth]{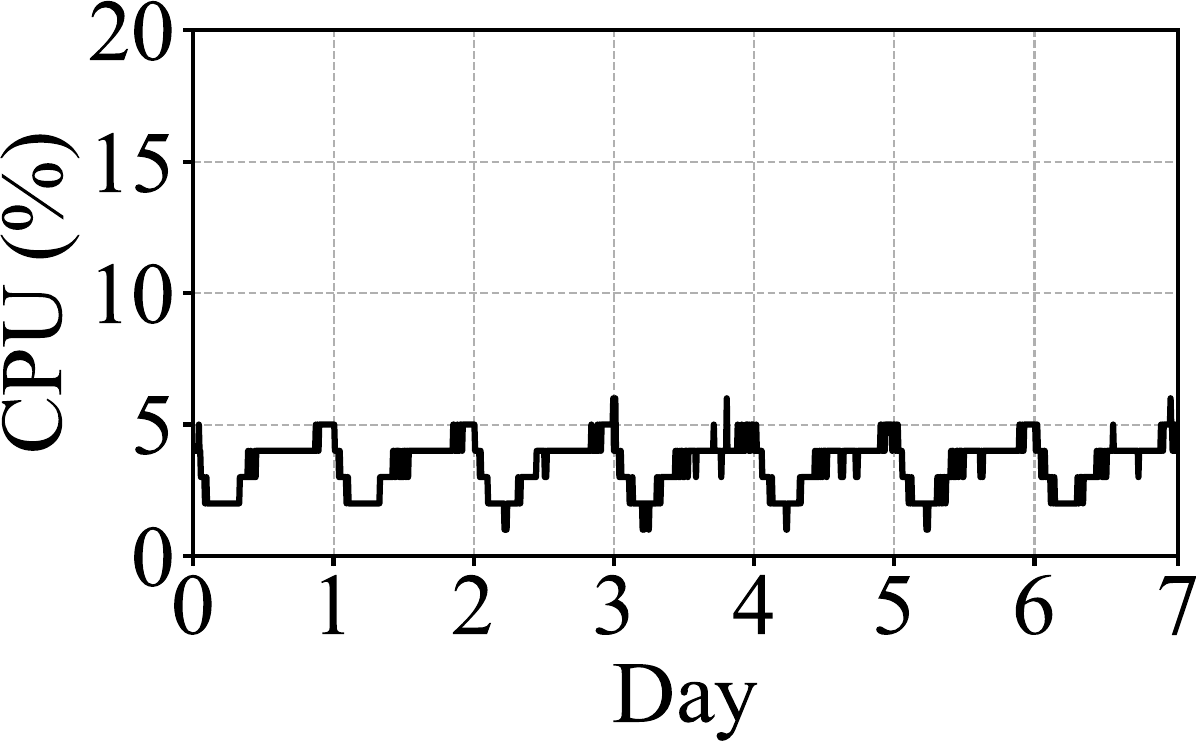}
    }
    \subfigure[City\#3]{
	\includegraphics[width=0.31\linewidth]{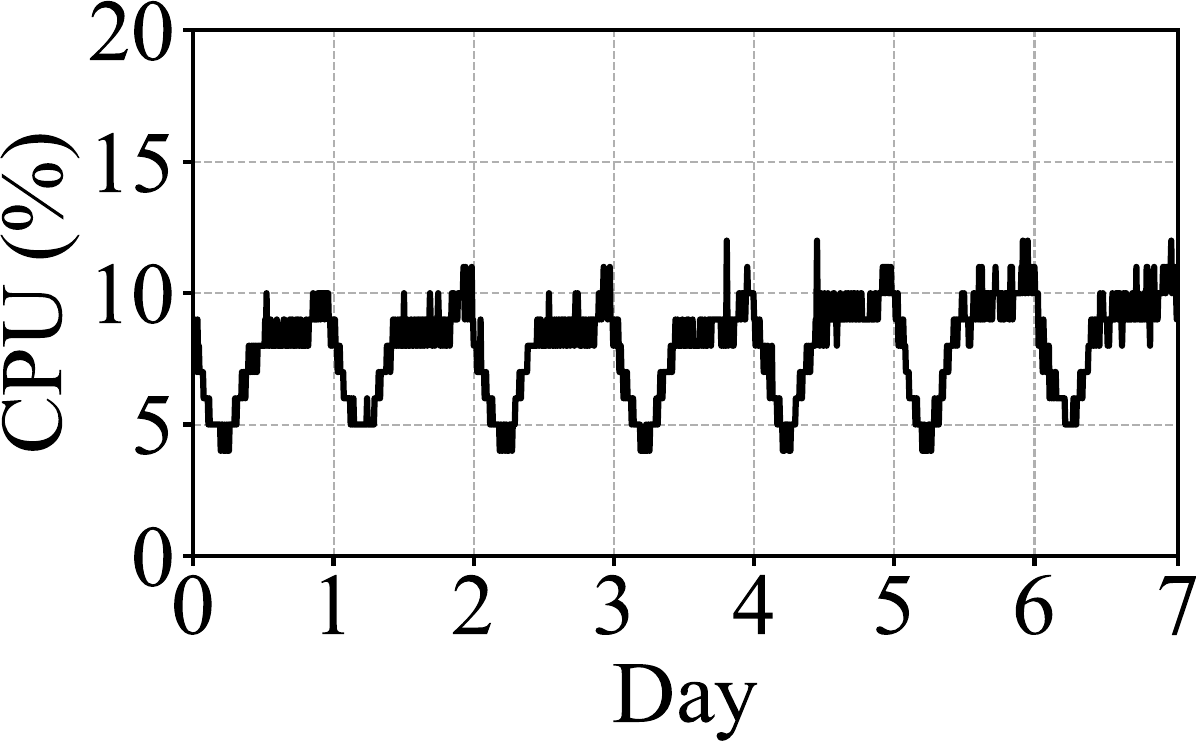}
    }
    \subfigure[City\#4]{
	\includegraphics[width=0.31\linewidth]{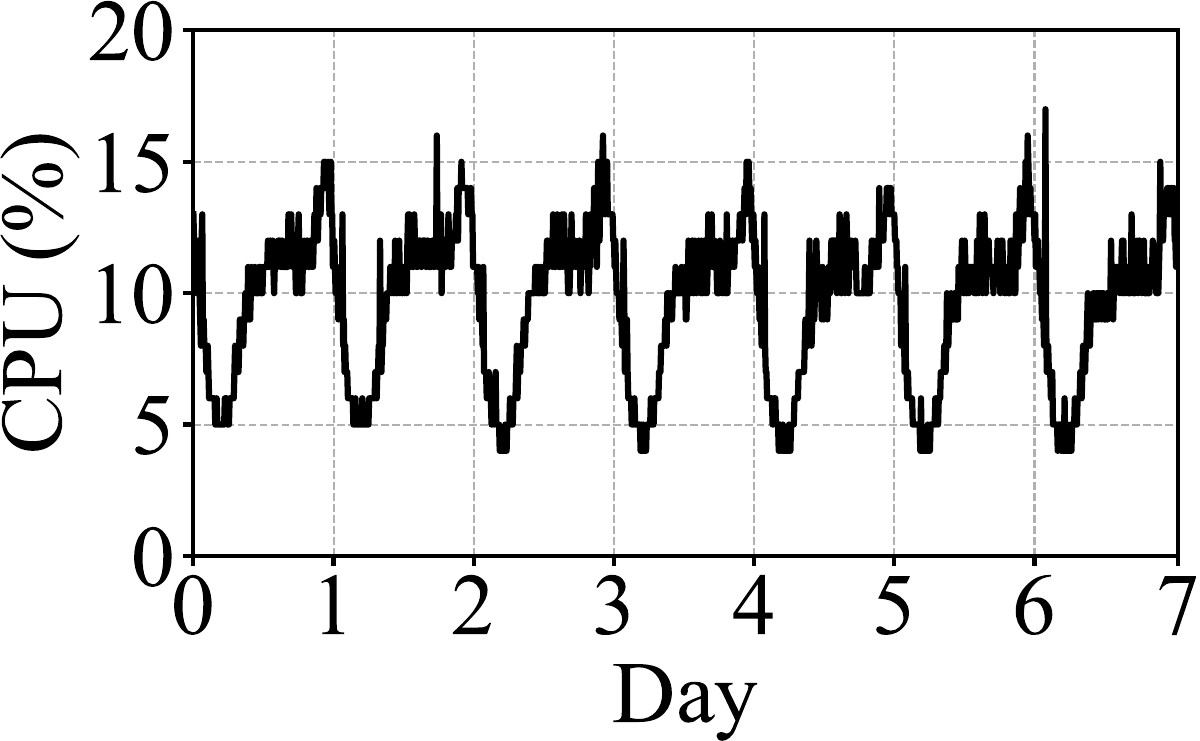}
    }
    \subfigure[City\#5]{
	\includegraphics[width=0.31\linewidth]{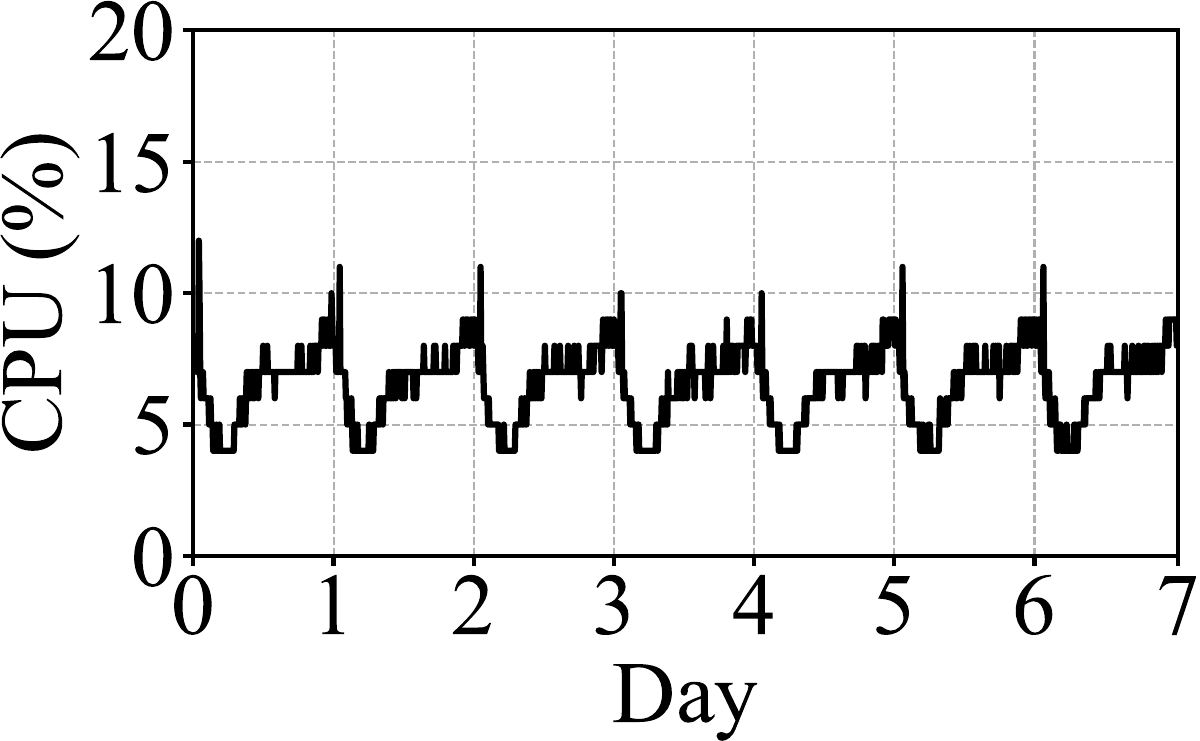}
    }
    \subfigure[City\#6]{
	\includegraphics[width=0.31\linewidth]{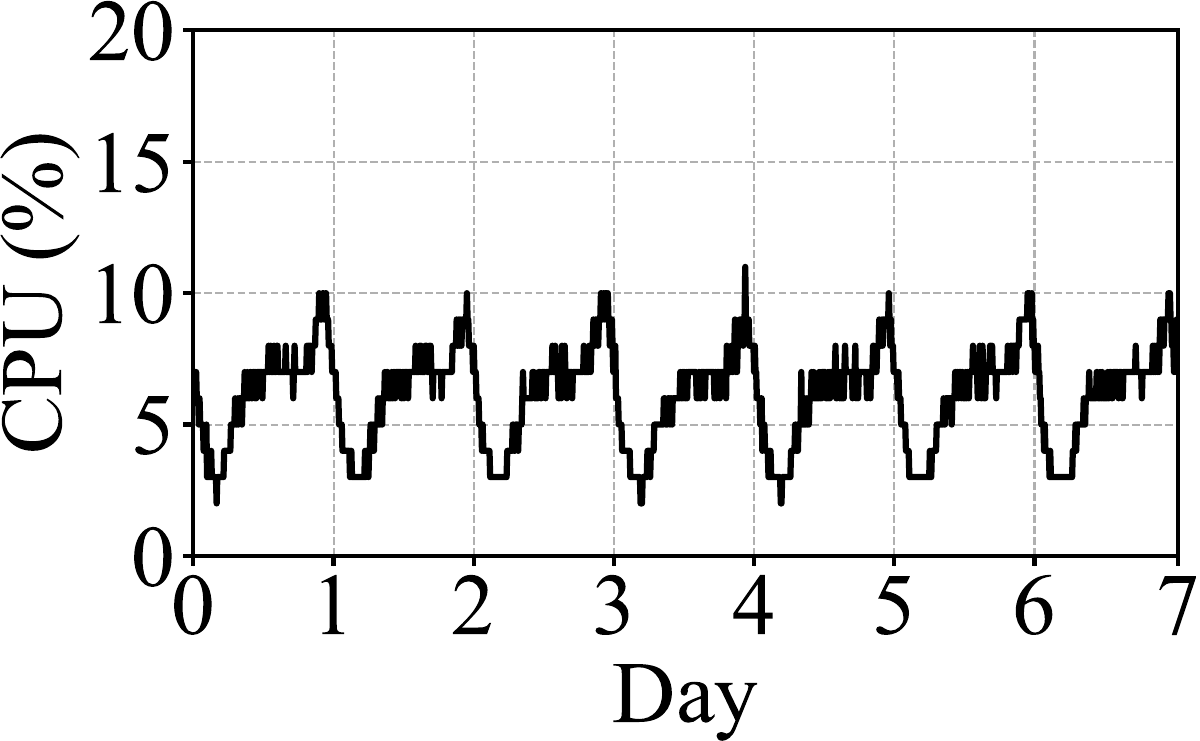}
    }
    \subfigure[City\#7]{
	\includegraphics[width=0.31\linewidth]{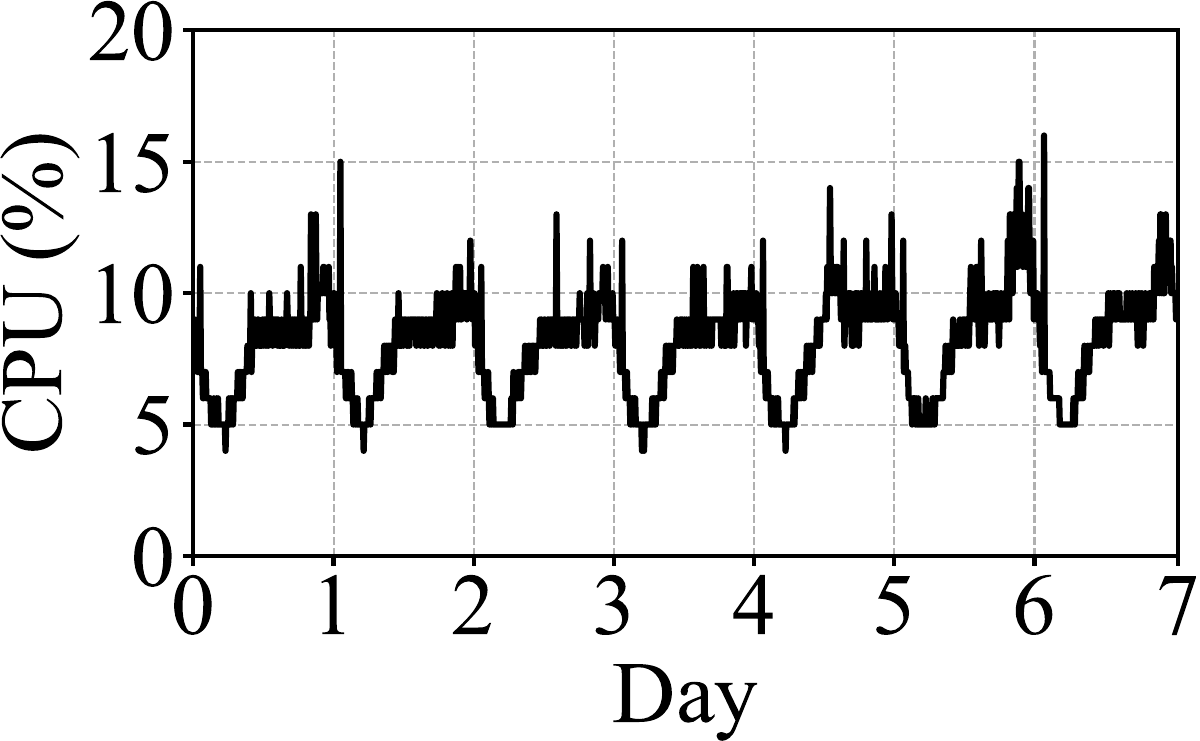}
    }
    \subfigure[City\#8]{
	\includegraphics[width=0.31\linewidth]{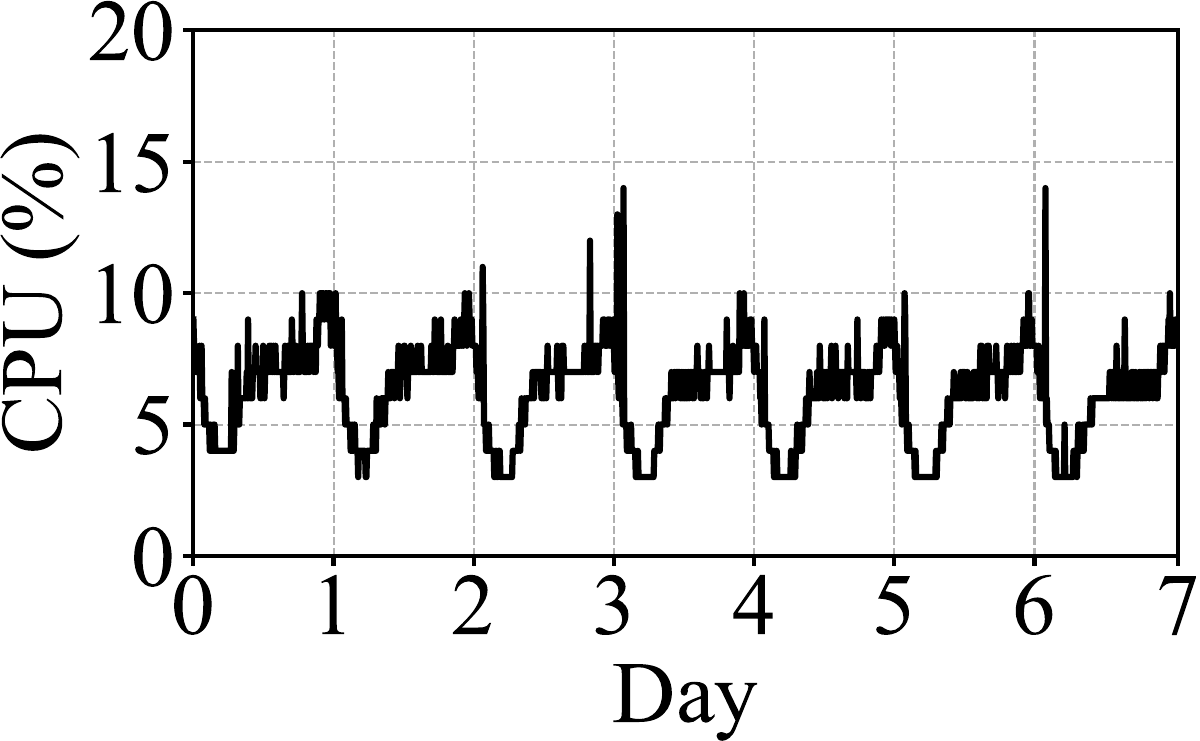}
    }
    \subfigure[City\#9]{
	\includegraphics[width=0.31\linewidth]{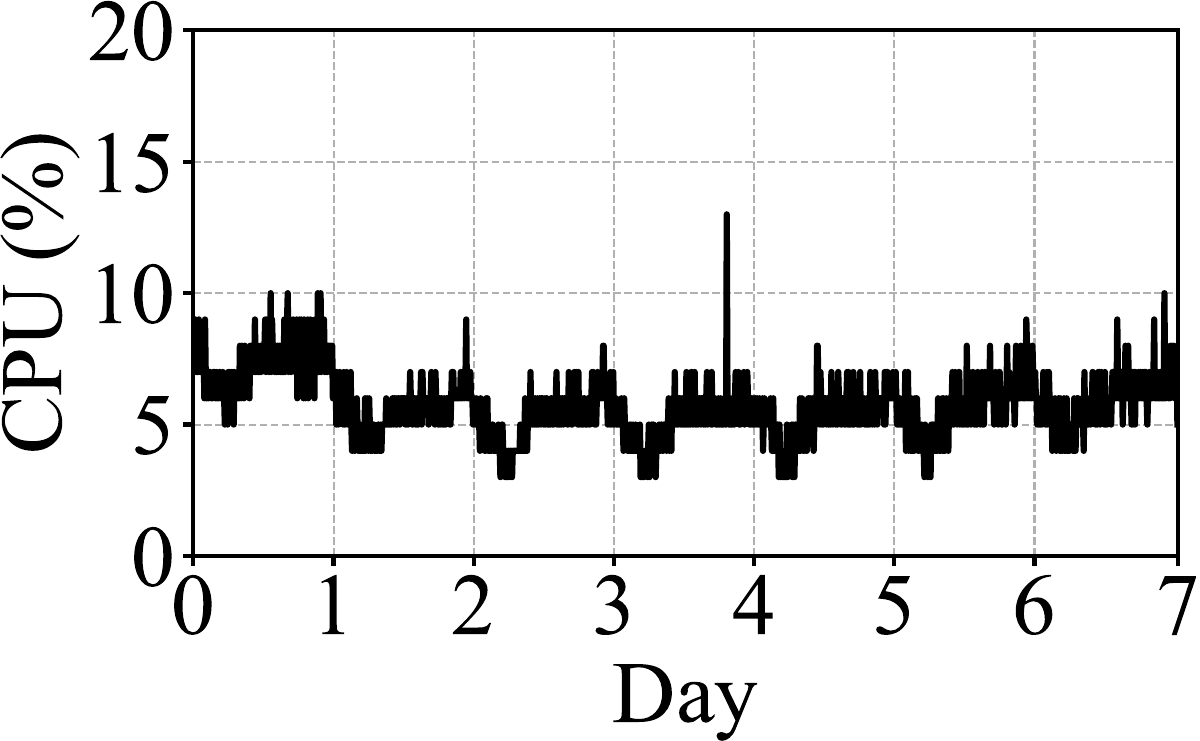}
    }
    \caption{The CPU utilization variation during a week on different edge nodes. City\#1-9 are the names of the cities where edge nodes are located.}
    \label{fig:cpu}
\end{figure*}

\begin{table}[t]
\caption{The latency monitored from \textit{Tencent} CDNs.}
\centering
\label{table:latency}
\begin{tabular}{l|c|c}
\hline
                & \textbf{Cache Hit} & \textbf{Cache Missing} \\ \hline
\textbf{P99 Latency (ms)}     & 9.70          & 218.06           \\ 
\textbf{P99.9 Latency (ms)}    &  15.31         & 283.71           \\
\textbf{Average Latency (ms)} & 1.90          & 231.07           \\ \hline
\end{tabular}
\end{table}

Content locality refers to the fact that many data items in disk storage share similar or even the same content~\cite{ren2010new}, which reflects the content redundancy phenomenon. This concept has been leveraged in the design of backup storage and data de-duplication strategies~\cite{morrey2006content,nachman2020goseed}. We believe that this concept can also be applied to CDNs to address cache-missing issues. Instead of fetching missing data from the data center, CDNs can generate it based on cached data with similar or related content. In this paper, we refer to the scenario of this specific cache miss as a "pseudo-missing" scenario. In this scenario, we propose a scheme called generative-hit and give the comprehensive definition in \S~\ref{sec:pseudo-miss}. This involves generating the requested data using local computing resources and relevant cached data. There are two reasons for adopting this scheme.

\textbf{Reason~\ding{202}: Generating data proves to be faster than fetching data in the majority of scenarios, and CDNs have sufficient computing resources to support data generation.} The advancement of technology for generating data has made it possible to control data generation efficiency precisely in most scenarios, typically within a range of 100 milliseconds~\cite{sajjadi2018frame}. By comparing the missing time displayed in Table~\ref{table:latency}, it can be concluded that the latency on CDNs can be reduced by returning locally generated data. Moreover, as shown in Figure~\ref{fig:cpu}, there are idle computing resources on CDNs. Therefore, generating data locally is feasible.

\begin{figure}[t]
    \centering
    \subfigure[Case A]{
	    \includegraphics[width=0.46\linewidth]{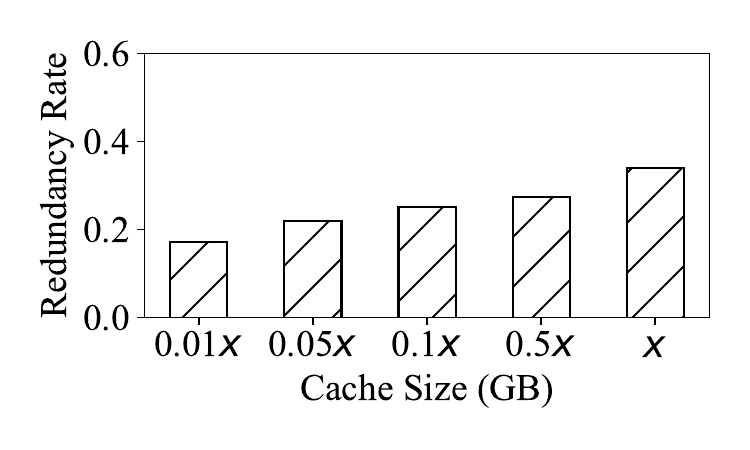}
	}
    \subfigure[Case B]{
	    \includegraphics[width=0.46\linewidth]{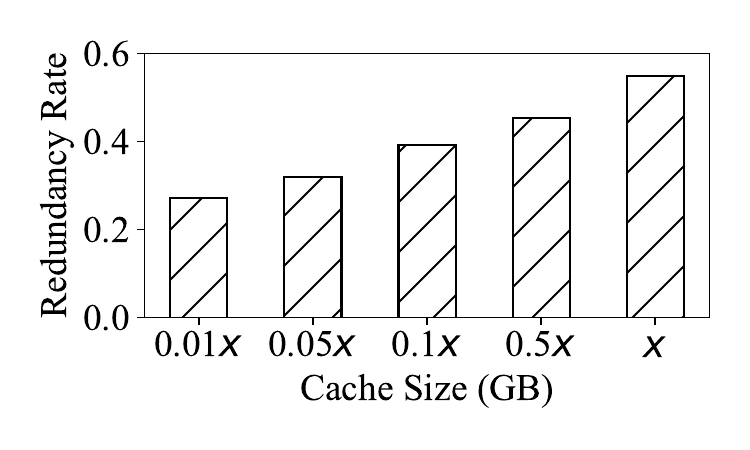}
    }
    \caption{Content redundancy rates of two CDN instances. $x=512$GB in Case A and $x=2048$GB in Case B.}
    \label{fig:rr}
\end{figure}

\textbf{Reason~\ding{203}: Generative-hit can effectively utilize content locality and enhance the content-carrying capacity of CDNs.} Content redundancy is often observed in CDNs. As shown in Figure~\ref{fig:rr}, we presented the distribution of data with identical content but varying specifications or formats on two CDNs, \ie, Case A~\cite{zhou2018demystifying} and Case B~\cite{liu2023slap}. Generative-hit can utilize content to generate data, which in turn reduces content redundancy to a reasonable level. This leads to an increase in cache utilization in terms of content and improves the quality of content-oriented services in CDNs.

Based on the above reasons, we propose a framework called \fname~that implements generative hit in CDNs. Using the block data as an illustration, once a user sends a request,~\fname~checks whether it hits via MD5. If not,~\fname~compares the URL of the request to the prefix of the filename of all files in the CDN to determine if it is a pseudo-missing scenario. In such a scenario,~\fname~handles the associated cached data and delivers the generated data to the user. Otherwise, the requested data is fetched and then written to the CDN. However, in cases where the pseudo-missing data are accessed frequently, the generating workload can become burdensome since the generated data is forbidden to be stored in the CDN. To tackle this issue, we introduce a learning model to predict the subsequent requests, asynchronously fetching the requested data when the above case occurs. 

Our contributions can be summarized as follows:

\ding{172} Based on content locality, we propose a generative-hit framework to handle pseudo-missing requests via data generation using cached data and computing resources on the edge nodes of CDNs.

\ding{173} Our prototype system has been deployed at \textit{Tencent}~\footnote{https://cloud.tencent.com/product/cdn}. The results show that the average access latency decreases by 56.04\% and the bandwidth used for fetching data (\ie, the back-to-COS bandwidth) reduces by 60.28\%.

\ding{174} Experimental results on the simulator show that running caching algorithms on~\fname~can further reduce the average access latency, back-to-COS bandwidth, and content redundancy.

\ding{175} We offer a new idea and solution to improve the service quality of CDNs using data generation. It will open the window for the application of large language models, which are good at generating data, on CDNs.

\section{Observations and Motivation}


\noindent\textbf{Observation}~\ding{202}~\textbf{: There are a large number of idle computing resources on CDNs.} We have surveyed CDN providers such as \textit{Akamai}\footnote{https://www.akamai.com/solutions/content-delivery-network}, \textit{Google}\footnote{https://cloud.google.com/cdn}, \textit{Alibaba}\footnote{https://www.alibabacloud.com}, \textit{Tencent}, \etc., and find that computing resources on CDNs are underutilized. Figure~\ref{fig:cpu} shows the average CPU utilization rates on different edge nodes of \textit{Tencent} CDNs. They are all lower than 20\%. This observation is consistent with the results of existing studies~\cite{khansoltani2022request,taleb2019cdn}. Note that as computing resources at the edge become abundant and CPU resources in the cloud become rare~\cite{taleb2019cdn,wang2021cost,xue2022scd2}, it becomes essential to offload computational functions to the edge nodes.

\noindent\textbf{Observation}~\ding{203}~\textbf{: "Generating" may be faster than "fetching".} We gathered tail latency and average latency for cache hits and missing during a month from \textit{Tencent} CDNs. As shown in Table~\ref{table:latency}, the time for missing is more than 100 times the time for hit. This time gap gives CDNs enough room to implement data generation and provide a generative hit. Based on our tests, we found that image format conversion or scaling usually takes around 40ms, while incremental transformations of data blocks have even shorter processing times, typically below 10ms.

\noindent\textbf{Observation}~\ding{204}~\textbf{: "Data generation" can be implemented in the security framework without copyright disputes.} Generating data on the edge node involves accessing the data in the cache, which may raise security concerns. While the topic of security is beyond the scope of this paper, we have made adaptations for the CDN environment by employing the Secure Content Delivery and Deduplication (SCD2) scheme~\cite{xue2022scd2}. In addition, the generated data only offers browsing services and cannot be stored in CDNs. Meanwhile, the generation tool is automatic and open source. Therefore, there are no copyright issues.

\noindent\textbf{Motivation.} Based on the observations above, we believe that the availability of computing resources, feasibility in terms of time, and security guarantee all support cache hits through data generation on CDNs. By fully utilizing the idle computing resources on the edge nodes, near-data processing can be achieved, which also fulfills the functional requirements of edge nodes in the design of new-generation distributed systems~\cite{taleb2019cdn}. As the pioneering study in this field, we will outline scenarios for utilizing generated data to facilitate hits on CDNs.

\section{Generative Hit}

\begin{figure*}[t]
    \centering
    \includegraphics[width=1\linewidth]{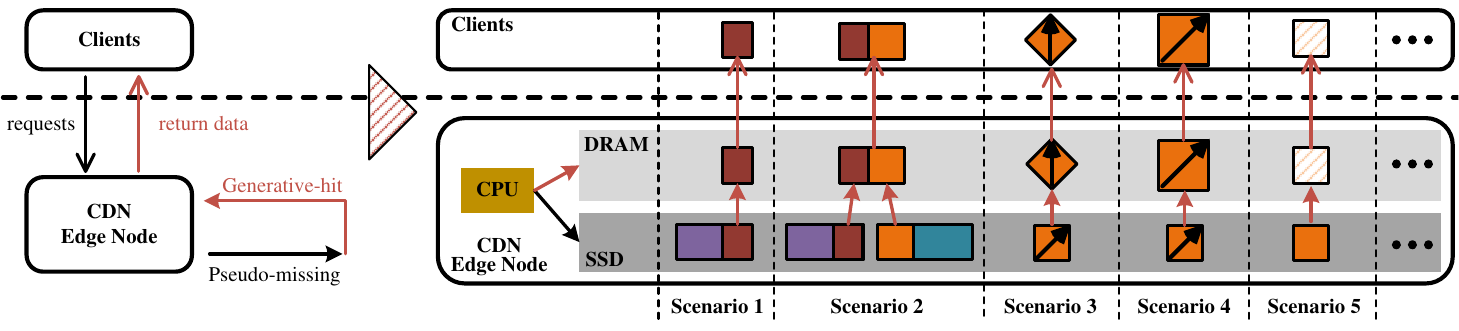}
    \caption{Several pseudo-missing scenarios and their generative-hit schemes.}
    \label{fig:pseudo_miss_case}
\end{figure*}

\subsection{Scenarios}
\label{sec:pseudo-miss}
The pseudo-missing scenario is that a user sends a request for data that is similar or related to the cached data in terms of content, where the requested data can be generated through cached data and local computing resources. Generative hit refers to the cache hit caused by using local computing resources and cached data on edge nodes to generate the requested data. When generating data is faster than fetching data, a pseudo-missing request can be responded to by a generative hit, resulting in decreased latency. We list five pseudo-missing scenarios and their generative-hit schemes in Figure~\ref{fig:pseudo_miss_case}.

\noindent $\bullet$ \textbf{Scenario 1 \& Scenario 2: The requested content is partially embedded in the cached data.} By using computing resources on the edge node, the corresponding part of the cached data is disassembled in DRAM and provided to the user. Similarly, the corresponding parts of data can also be combined in DRAM and then the results are pushed to the user.

\noindent $\bullet$ \textbf{Scenario 3 \& Scenario 4: The content embodied in cached data is identical to the requested content, but these data have different forms (\eg, shape, size, format,~\etc.).} The copy of the corresponding cached data will be processed (\eg, rotating, scaling, converting,~\etc.) to fit the user's request and sent to the user.

\noindent $\bullet$ \textbf{Scenario 5: The content embodied in cached data is similar to the requested content.} The copy of the corresponding cached data will be revised (\eg, enhancing, transcoding,~\etc.) to fit the user's request and sent to the user.

The data generation for the above scenarios draws on the content locality characteristic. To minimize content redundancy and reduce SSD write times, the generated data is only used for responding to the user's \mname~requests and is not stored on local SSDs.

\begin{figure}[t]
    \centering
    \includegraphics[width=1\linewidth]{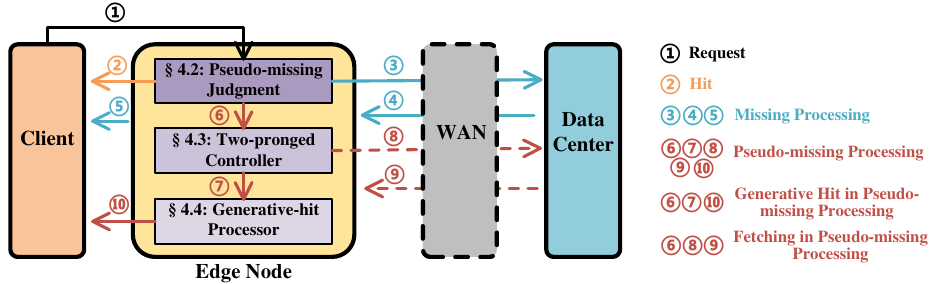}
    \caption{The workflow of hit, missing, and pseudo-missing in the architecture of~\fname.}
    \label{fig:pseudo_miss}
\end{figure}

\subsection{Rationale}
\label{sec:ipf}

Figure~\ref{fig:pseudo_miss} represents the workflow of hit, missing, and pseudo-missing. We use the order number in Figure~\ref{fig:pseudo_miss} to represent the latency resulting from each corresponding operation/behavior. 
We define the latency for a pseudo-missing request and a missing request as $\mathcal{T}^*$=\ding{192}+\ding{197}+\ding{198}+\ding{201} and $\mathcal{T}$=\ding{192}+\ding{194}+\ding{195}+\ding{196}, respectively. Given $n$ requests for the data with the same content and their access times $t_1, t_2, ..., t_n$, we discuss the average access latency of these requests in the missing and pseudo-missing scenarios. Note that we assume that \ding{192}+\ding{193}=0 since the latency of a cache hit is much smaller than that of a missing request.

$\bullet$ \textbf{Missing:} According to the traditional caching policy, if a cache missing occurs at $t_1$, the latency of this access is $\mathcal{T}$. At $t_2$, if $t_2-t_1>\mathcal{T}$, the second access will hit with a latency of 0. If ensuing requests at $t_3, ..., t_n$ are hit before the data with the requested content is evicted, the optimal average latency of these $n$ requests is $\frac{\mathcal{T}}{n}$.

$\bullet$ \textbf{Pseudo-missing:} Assuming that a pseudo-missing request occurs at $t_1$, resulting in a latency of $\mathcal{T}^*$. At $t_2$, the pseudo-missing will occur again and yield a latency of $\mathcal{T}^*$. The same latency will be yielded at $t_3, ..., t_n$. As a result, the average latency of these $n$ requests is $\mathcal{T}^*$.

$\bigstar$ \textbf{Analysis:} When $\frac{\mathcal{T}}{n}>\mathcal{T}^*$, the latency is lowered because of the new policy designed for pseudo-missing. Note that as $n$ increases, it is possible for $n\times \mathcal{T}^*>\mathcal{T}$. We attribute it to giving up on writing SSDs for generated data. However, it is conducive to the reduction of data redundancy and the preservation of content diversity. To achieve a better balance between the two, we have developed further optimization strategies.

$\bullet$ \textbf{Further optimization for pseudo-missing:} To further reduce the average access latency caused by pseudo-missing without excessively reducing content diversity, we adopt a two-pronged strategy. As \ding{199} and \ding{200} shown in Figure~\ref{fig:pseudo_miss}, while using the generated data to respond to user requests promptly, the edge node still fetches data and writes it to SSDs. Assume that at the $p$-th request, the two-pronged strategy achieves the above operation before the arrival of the $q$-th request. The average latency of pseudo-missing can be optimized to $\frac{q-1}{n}\times\mathcal{T}^*$ because of $t_{q-1}-t_p<\mathcal{T}<t_q-t_p$. Furthermore, since the original data obtained asynchronously can be provided for a hit in the next request, there is no need to write the generated data to SSDs.

$\bigstar$ \textbf{Analysis:} Although the average latency can be smaller as $q$ decreases, this strategy is constrained by the actual situation. For example, if there is only one access, fetching is meaningless. Even if $q>1$ but the latency for fetching data is greater than that of pseudo-missing processing, it still causes performance degradation. Consequently, to prevent meaningless fetching, we need to establish a decision model that determines whether the strategy should be executed. In addition, the model should take into account the availability of idle computing resources. The implementation of the decision model is described in \S~\ref{sec:two-pronged}.

\section{C\MakeLowercase{o}G\MakeLowercase{en}T Design}
Based on the above prefabrication and analysis, we propose a content-oriented generative-hit framework (\fname) that can be deployed through a simple reconfiguration of CDNs.

\subsection{Overall Architecture}

We illustrate the overall architecture of \fname~in Figure~\ref{fig:pseudo_miss}. It involves four participants: the client, the CDN edge node, the wide area network (WAN), and the data center. Our work focuses on enhancing the edge node. Specifically, we introduce three modules: the pseudo-missing judgment module (see \S~\ref{sec:judge_module}), the two-pronged controller (see \S~\ref{sec:two-pronged}), and the generative-hit processor (see \S~\ref{sec:epmp}). The pseudo-missing judgment module checks user requests to determine whether a request belongs to hit, missing, or pseudo-missing. The two-pronged controller leverages a decision tree model to predict subsequent access and determine whether to fetch and write data. 
The generative-hit processor generates data and provides the generated hit by utilizing local computing resources to process cached data.

\subsection{Pseudo-missing Judgment Module}
\label{sec:judge_module}

This module takes into account the time spent generating data and idle CPU resources. When the time taken to generate data exceeds the time taken to fetch data, or when there are insufficient CPU resources, the module will shield pseudo-missing operations and handle the request using traditional patterns. We develop a set of global identifiers in the judgment function (\ie, \texttt{get\_metadat(URLReq req, TNode \&node)}) to represent the hit, missing, and pseudo-missing scenarios. Each identifier for block data comprises the file name (\ie, \texttt{req.key}) along with specific parameters (\ie, \texttt{req.par}). If the file name in the URL request does not match any of the cached files, the requested data is considered missing. The requested data will be fetched from the data center, written to the edge node, and then sent to the client. These behaviors correspond to \ding{194}, \ding{195}, and \ding{196} shown in Figure~\ref{fig:pseudo_miss}, respectively. For requests where the file name matches but the related parameters differ, they are processed as pseudo-missing. The behavior \ding{197} means the start of pseudo-missing processing. If both the file name and parameters match exactly, it is considered a cache hit. The data from the hit will be sent to the client according to the behavior \ding{193}.

\subsection{Two-pronged Controller}
\label{sec:two-pronged}

The two-pronged controller performs two functions simultaneously. On the one hand, it issues instructions to generate the requested data. On the other hand, it asynchronously fetches data from the data center and writes it to the edge nodes when it predicts that the pseudo-missing data will be frequently accessed. As the prediction component of the controller, the decision model should accurately predict the reusability of requested files. In the implementation, we employ a decision tree model to achieve prediction, referring to~\cite{wang2018efficient}. To ensure effective predictions, we extract several features, including file type, file size, age of file, recency, and frequency, to learn the prediction model. 

\begin{figure}[t]
    \centering
    \includegraphics[width=0.9\linewidth]{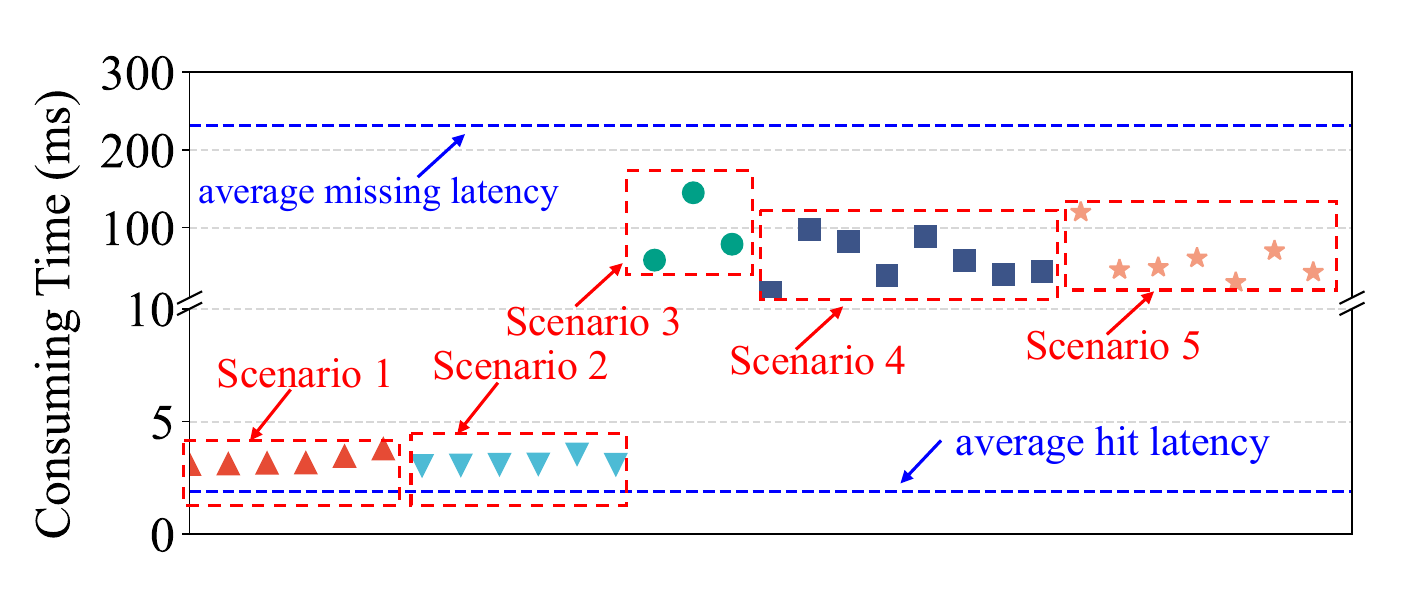}
    \caption{The latency of generative hits in different pseudo-missing scenarios.}
    \label{fig:pm_latency}
\end{figure}

\subsection{Extensible Generative-hit Processor}
\label{sec:epmp}
In this module, we focus on generating the requested data for the scenarios mentioned in $\S~\ref{sec:pseudo-miss}$. According to the available open-source dataset, we initially combine the processing functions for image data and block data. To facilitate the extension for handling other types of data (\eg, text, video,~\etc.), we record the format and modality of the data in \texttt{metadata}. For multimedia data, we will additionally record the 128-bit similarity hash codes~\cite{liu2018deep} that represent the content of the data. According to the extended \texttt{metadata}, all types of cached data can be evaluated to determine their eligibility for participating in the pseudo-missing processing. If eligible, they will be sent to the appropriate algorithm model for data generation.

$\bullet$ \textbf{For image data.} We introduced a third-party open-source image processing library\footnote{https://github.com/nackily/imglib} to achieve image generation in Scenario 3, Scenario 4, and Scenario 5 mentioned in $\S~\ref{sec:pseudo-miss}$. This library is built on image-processing techniques developed in C++. To achieve on-the-fly processing, we prefer to execute efficient algorithms. If higher content quality is required, more complex algorithms can be selected instead.

$\bullet$ \textbf{For block data.} In Scenario 1 and Scenario 2 mentioned in $\S~\ref{sec:pseudo-miss}$, the required data blocks are separated according to the parameters given by the user URL. If a merge operation is required, data is successively read from the source files into the merge file after the files that need to be merged have been sorted and an empty merge file has been created.

As shown in Figure~\ref{fig:pm_latency}, we measured the time for generative hits in five scenarios mentioned in $\S~\ref{sec:pseudo-miss}$. For block data, we measured the generative-hit time in Scenario 1 and Scenario 2 using different block sizes (ranging from 4KB to 1MB). It can be observed that the generative-hit time is around 1ms. In addition, the generative-hit time in Scenario 3, Scenario 4, and Scenario 5 for image data consistently remains below 150ms. Referring to the latency yielded by traditional processing shown in Table~\ref{table:latency}, we conclude that generative hits can effectively reduce latency.

\section{Evaluation}
We have implemented a prototype within \textit{Tencent} CDNs (\textit{PicCloud}) and tests~\fname~on the simulator~\cite{atre2020caching} based on two open-source traces~\cite{zhou2018demystifying,liu2023slap}. To determine the cached data involved in the generative hit for block data, we check the prefix of the file names. For image data, we check the Hamming distances between similarity hash codes calculated by the DSTH model~\cite{liu2018deep}. When \fname~was deployed, we maintained the cache replacement policy (LRU) used in the original system. The metrics we focus on are access latency and back-to-COS bandwidth.

\begin{figure*}[t]
    \centering
    \includegraphics[width=1\linewidth]{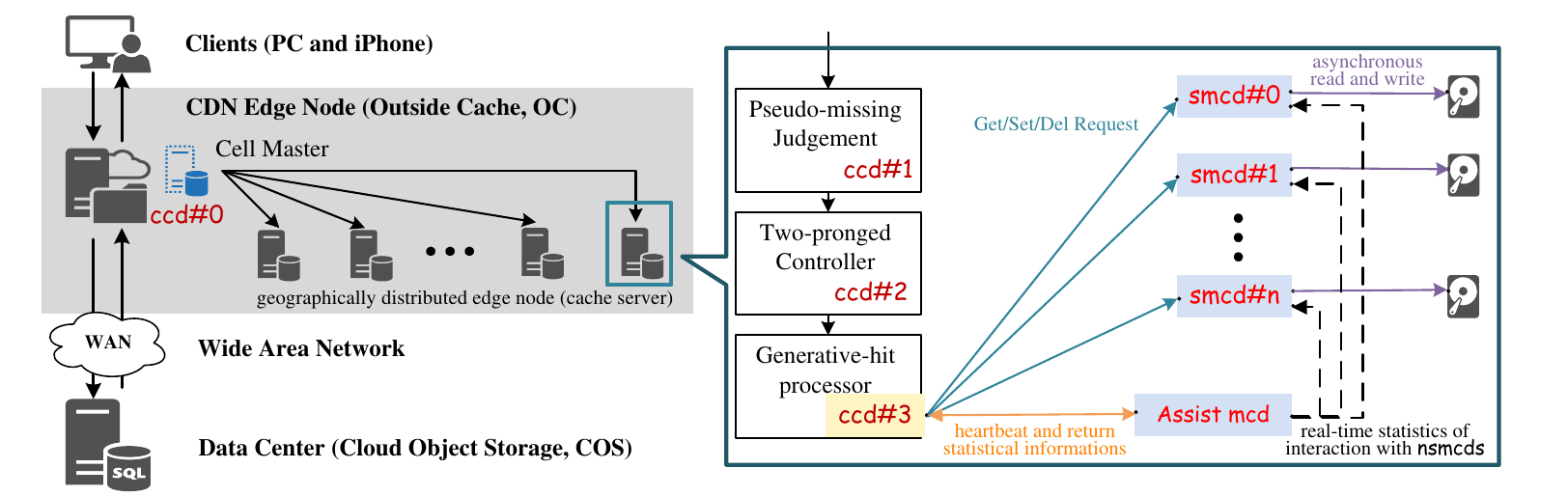}
    \caption{Deploy \fname~on \textit{Tencent} CDN (\textit{PicCloud}).}
    \label{fig:ccdn}
\end{figure*}

\begin{figure*}[t]
    \centering
    \subfigure[Average latency]{
	    \includegraphics[width=0.46\linewidth]{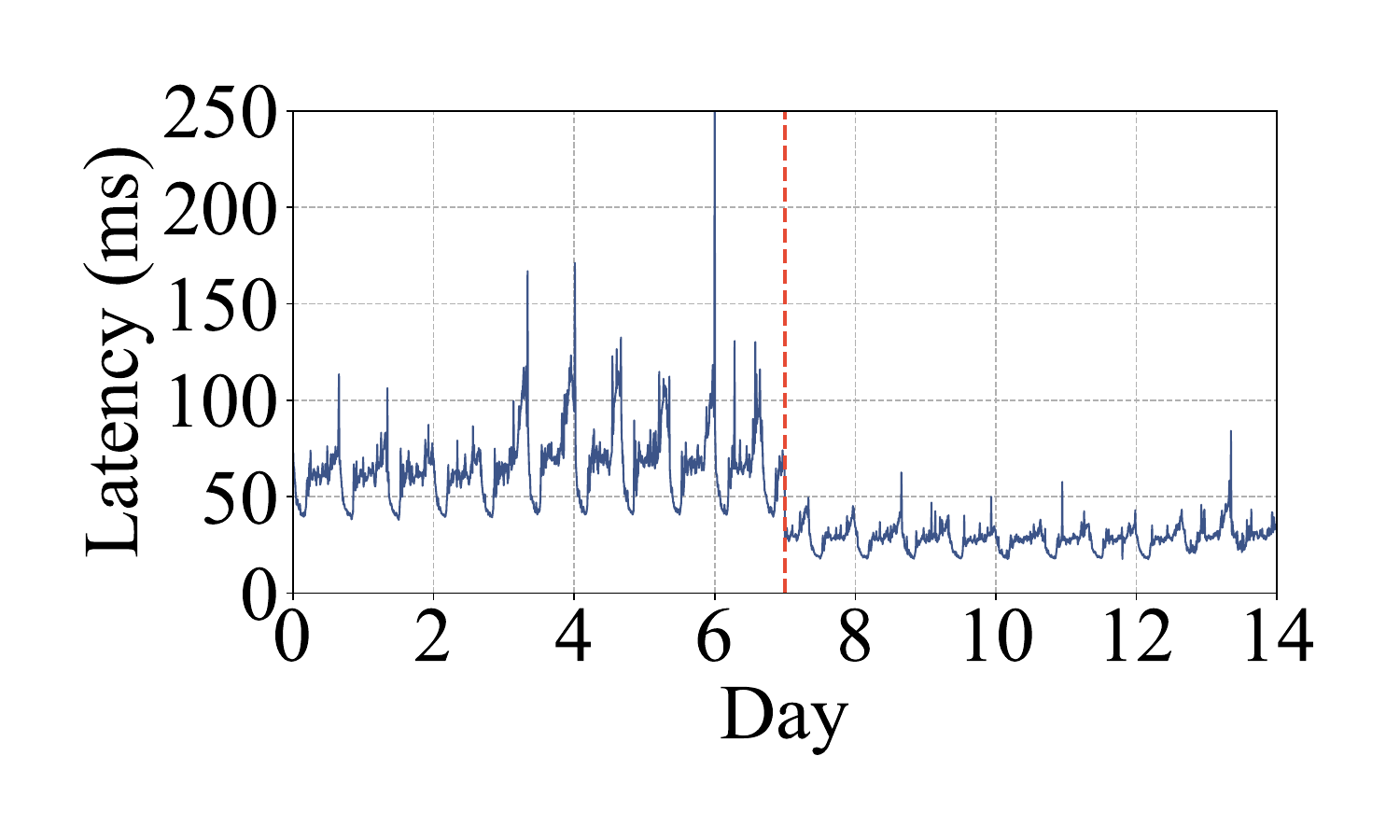}
	    \label{fig:real-latency}
	}
    \subfigure[Back-to-COS bandwidth]{
	    \includegraphics[width=0.46\linewidth]{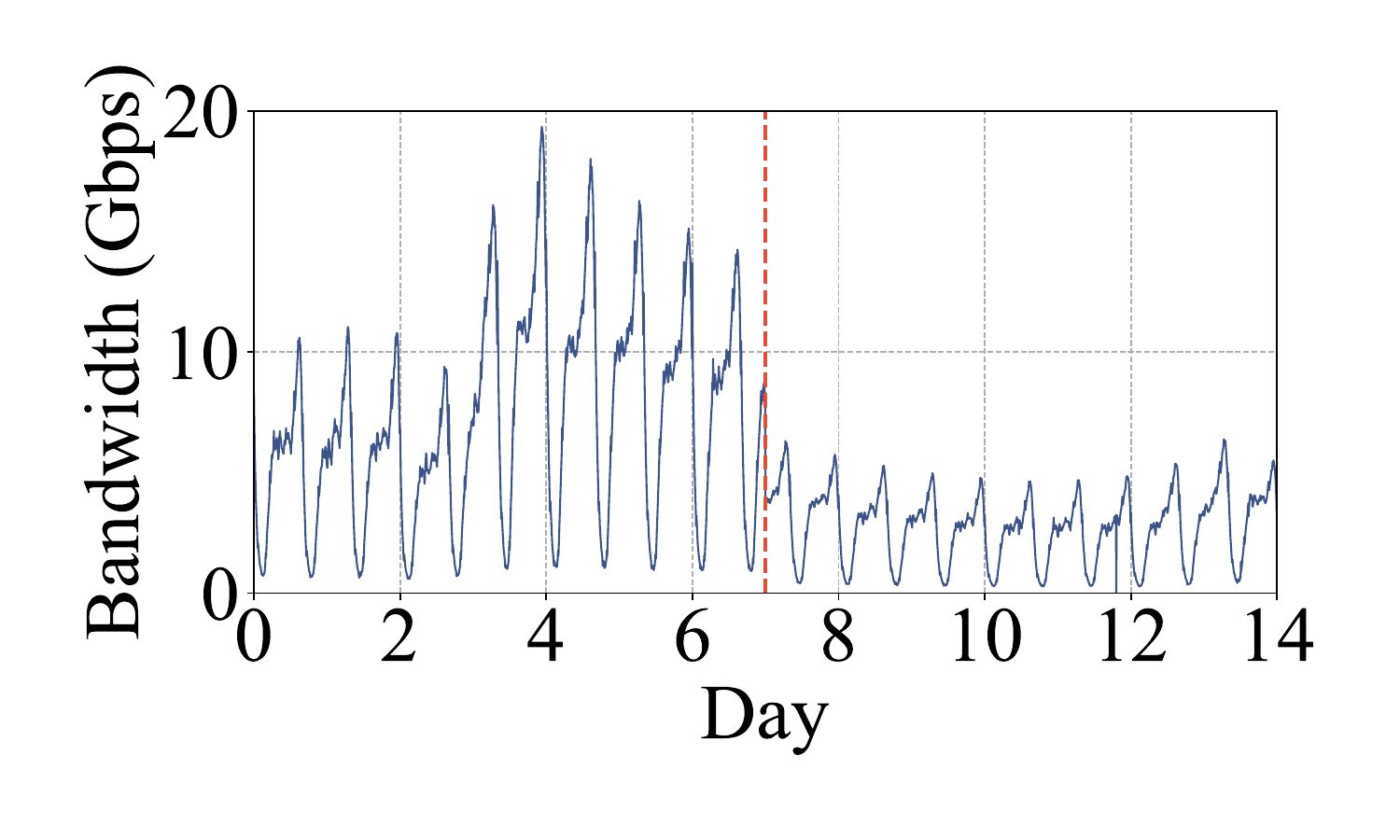}
	    \label{fig:real-bandwidth}
    }
    \subfigure[CPU utilization]{
	    \includegraphics[width=0.46\linewidth]{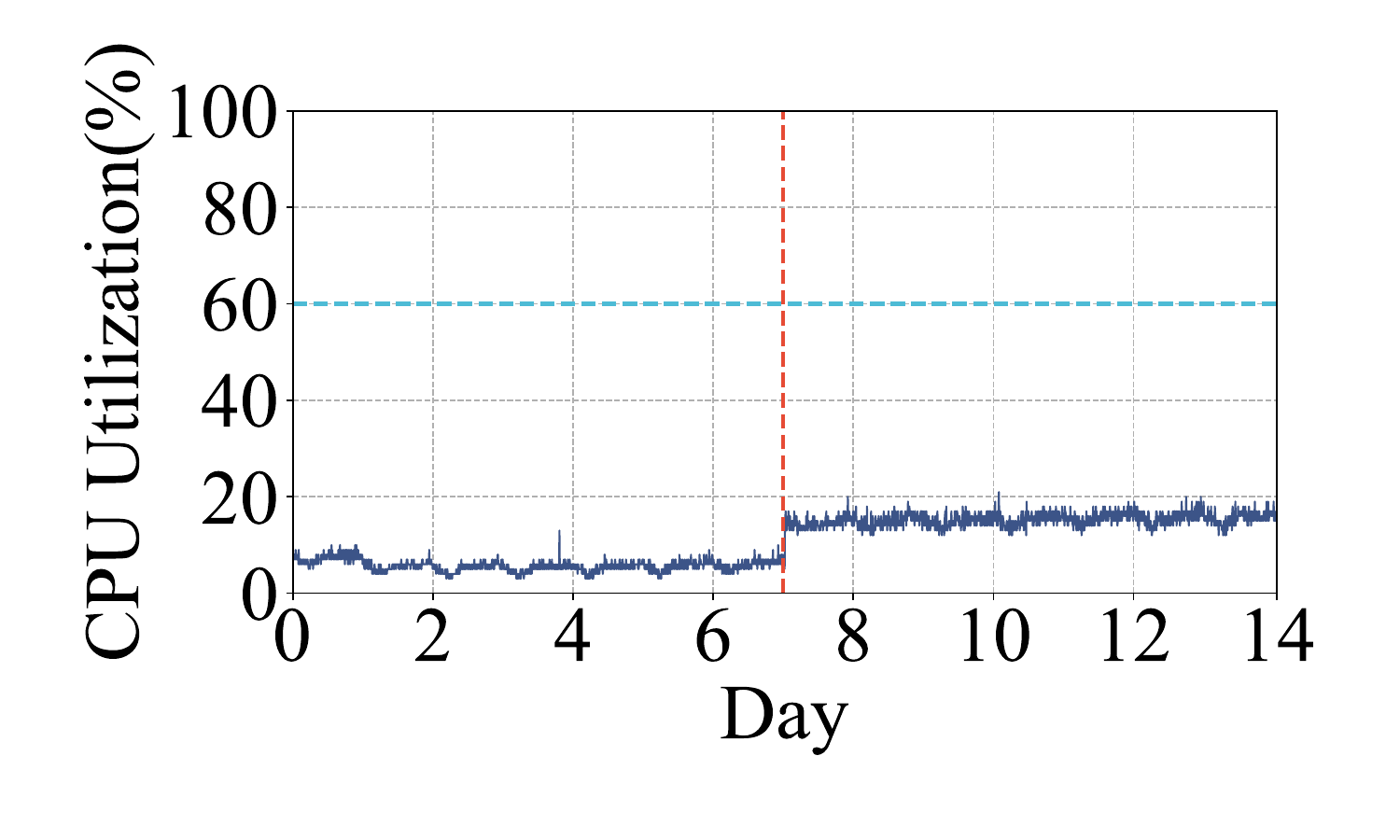}
	    \label{fig:real-cpu}
	}
    \subfigure[Memory]{
	    \includegraphics[width=0.46\linewidth]{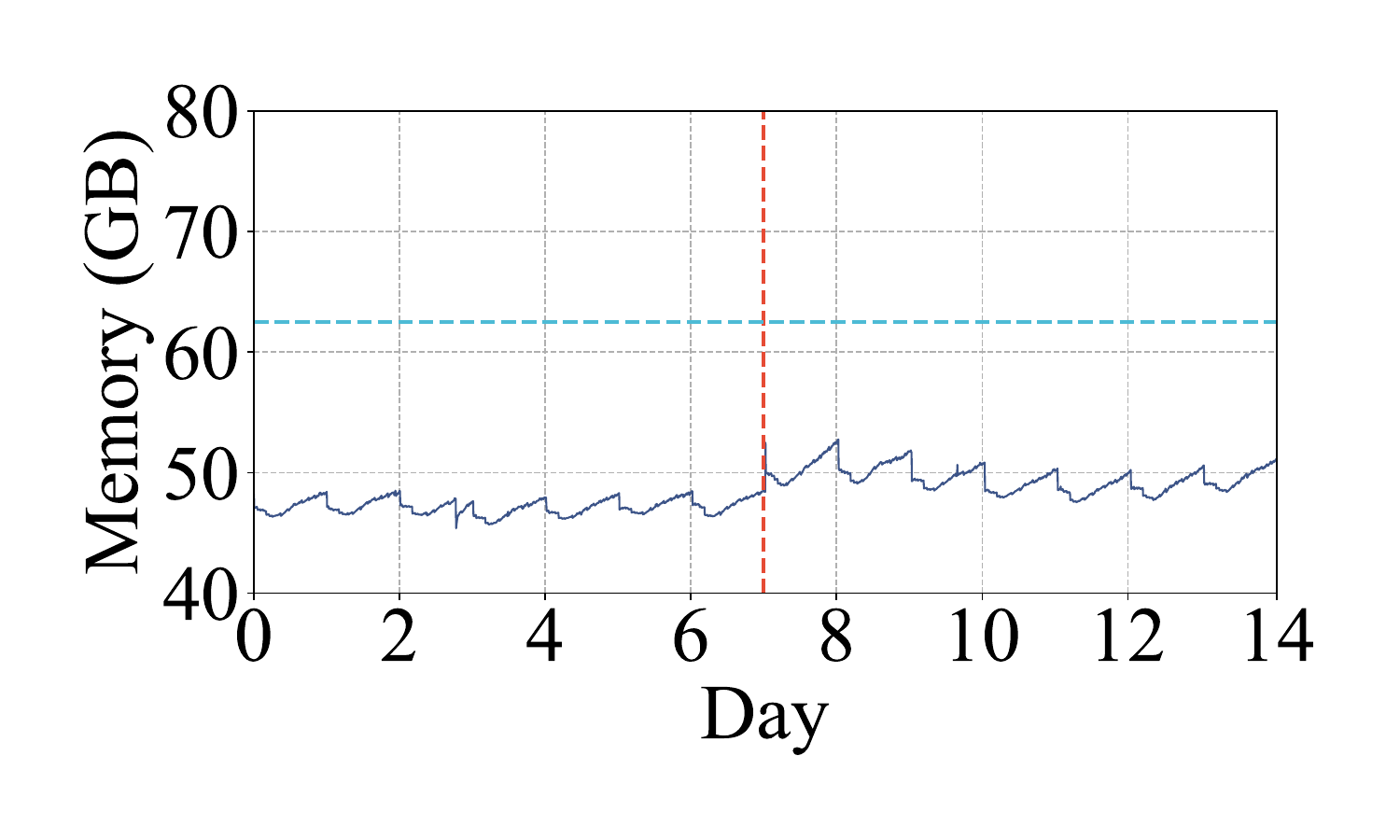}
	    \label{fig:real-memory}
    }
    \caption{Comparison of average latency and back-to-COS bandwidth, CPU utilization, and memory usage before and after deploying \fname. The red vertical dashed line indicates the date when \fname~was deployed. The blue horizontal dashed lines in (c) and (d) are the default upper limits (\ie, 60\% and 64GB) for CPU utilization and memory usage, respectively, which are used to reserve buffers to deal with sudden traffic surges.}
    \label{fig:real}
\end{figure*}

\subsection{Deployment of prototype}
As shown in Figure~\ref{fig:ccdn}, \textit{PicCloud} is a multithreaded, event-based CDN that configures memory and SSDs. Note that we placed the indexes in the memory rather than SSDs to prolong the lifespan of SSDs. \textit{PicCloud} consists primarily of two modules: \textit{Cell Master} and \textit{Cache Server}. As a resource management module, the \textit{Cell Master} manages the routing information of the entire system. It runs a process independently (\ie, \texttt{ccd\#0}), periodically checks the running status of each \textit{cache server}, and receives statistics reported by each \textit{cache server}. The \textit{Cache Servers} are geographically distributed edge nodes, used for the storage and processing of data. Its prototype is a storage node based on MCP++, a multi-\texttt{ccd}/multi-\texttt{smcd} process model, raw disks, inodes, and asynchronous disk I/O technologies.

\fname~is distributed on each edge node that runs three processes. The three processes manage the pseudo-missing judgment module (\ie, \texttt{ccd\#1}), the two-pronged controller module (\ie, \texttt{ccd\#2}), and the generative-hit processor module (\ie, \texttt{ccd\#3}), respectively. Each thread \texttt{smcd} corresponds to a physical disk and manages the corresponding disk in a single thread. The thread \texttt{Assist mcd} process is responsible for managing \texttt{n} \texttt{smcd}s, including collecting heartbeat information and statistics on the latency of get/set/del operation, disk data elimination information, workload,~\etc.

\subsection{Improvement and overhead on the real production system}
\noindent\textbf{Improvement.} As shown in Figure~\ref{fig:real-latency} and Figure~\ref{fig:real-bandwidth}, we measure the performance changes in the average access latency and back-to-COS traffic in \textit{PicCloud}. After updating the system architecture, the average access latency drops 56.04\%, and the average back-to-COS bandwidth drops 60.28\%. We also obtain the changes of the P99 and P99.9 latency after \fname~deployment from the monitoring system. As shown in Table~\ref{table:tail_latency}, the P99 latency drops by 74.91ms and the P99.9 latency drops by 108.56ms. These results confirm that \fname~can bring performance improvements.

\begin{table}[t]
\caption{The tail latency monitored from \textit{PicCloud}.}
\centering
\label{table:tail_latency}
\scalebox{1}{
\begin{tabular}{l|c|c}
\hline
                & \textbf{Original System} & \textbf{\fname~Deployed} \\ \hline
\textbf{P99 Latency (ms)}     & 118.37 & 43.46           \\ 
\textbf{P99.9 Latency (ms)}    &  171.23         & 62.67           \\ \hline
\end{tabular}
}
\end{table}

\noindent\textbf{Overhead.} To visualize the changes in physical machine metrics after the deployment of \fname, we depict the changes in CPU utilization and memory usage in Figure~\ref{fig:real-cpu} and Figure~\ref{fig:real-memory}, respectively. After deploying \fname, the CPU utilization increases from 5.73\% to 15.23\%, and the memory usage increases from 47.16GB to 49.49GB. As a result, we believe that trading idle resources for an increase in cache performance is worth it, albeit in additional CPU and memory resources required for data processing.

\subsection{Superiority, enhancement effect, and content redundancy tested on the simulator}

\begin{figure}[t]
    \centering
    \subfigure[Average latency]{
	    \includegraphics[width=0.46\linewidth]{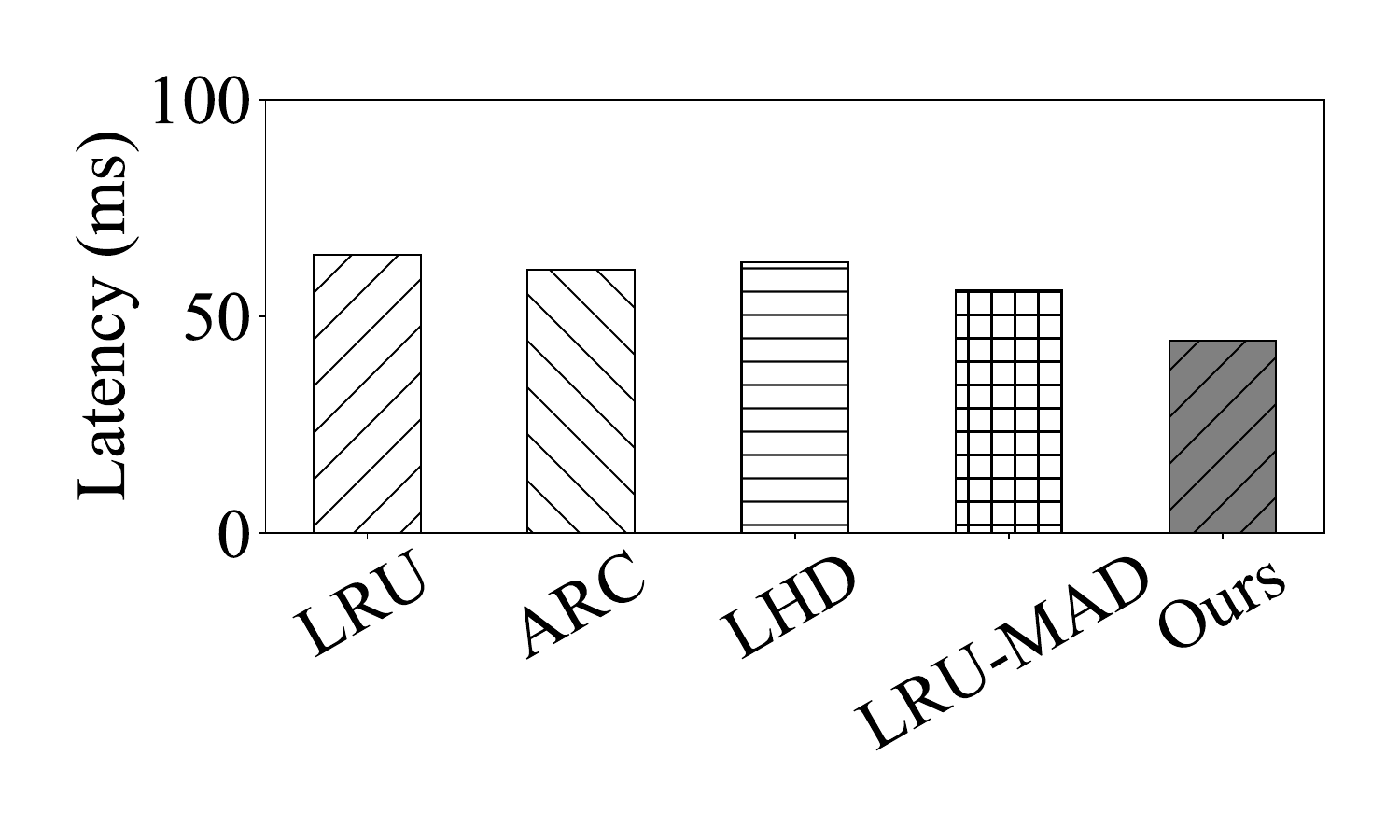}
	    \label{fig:latency1}
	}
    \subfigure[Back-to-COS bandwidth]{
	    \includegraphics[width=0.46\linewidth]{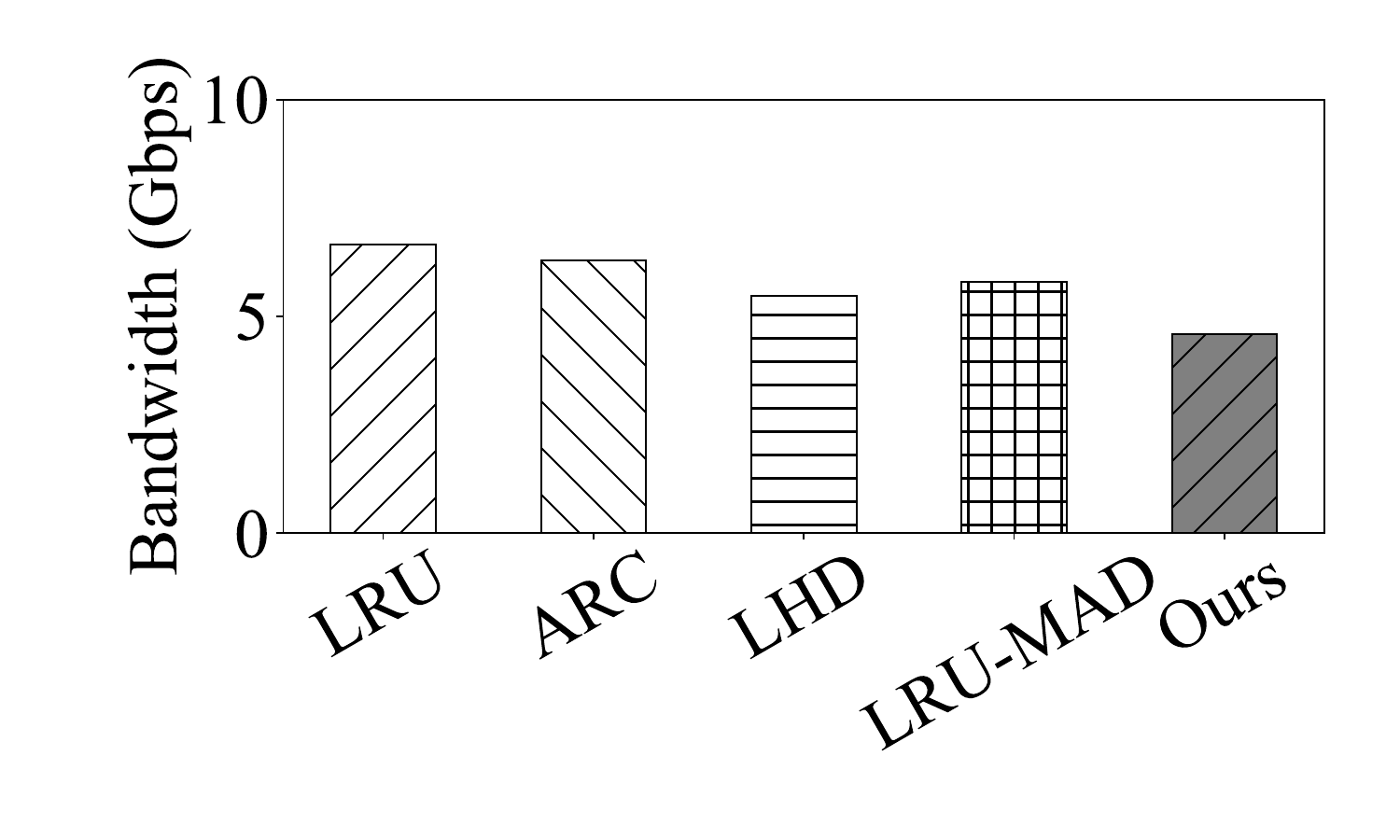}
	    \label{fig:bandwidth1}
    }
    \caption{Compared LRU on~\fname~ (\ie, Ours) with other algorithms on the original system in terms of average access latency and back-to-COS bandwidth.}
    \label{fig:real}
\end{figure}


We use the modified simulator used in LRU-MAD~\cite{atre2020caching} to test the metrics yielded by the representative caching algorithms. The algorithms include LRU, ARC~\cite{megiddo2003arc}, LHD~\cite{beckmann2018lhd}, and LRU-MAD~\cite{atre2020caching}, where LRU-MAD is a latency-sensitive algorithm. For Case A and Case B, we handle pseudo-missing in Scenario 3 and Scenario 4 as well as in Scenario 1 and Scenario 2, respectively. 

\noindent\textbf{For superiority}, we compare running LRU on~\fname~with running other caching algorithms on the original system. As shown in Figure~\ref{fig:latency1}, LRU on~\fname~yields an average access latency of 19.95ms, which is lower than those of other algorithms on the original system. 
In terms of back-to-COS bandwidth shown in Figure~\ref{fig:bandwidth1}, 
our result is 1.21Gbps lower than the optimal result of other algorithms. 

\begin{figure}[t]
    \centering
    \subfigure[Average latency]{
	    \includegraphics[width=0.46\linewidth]{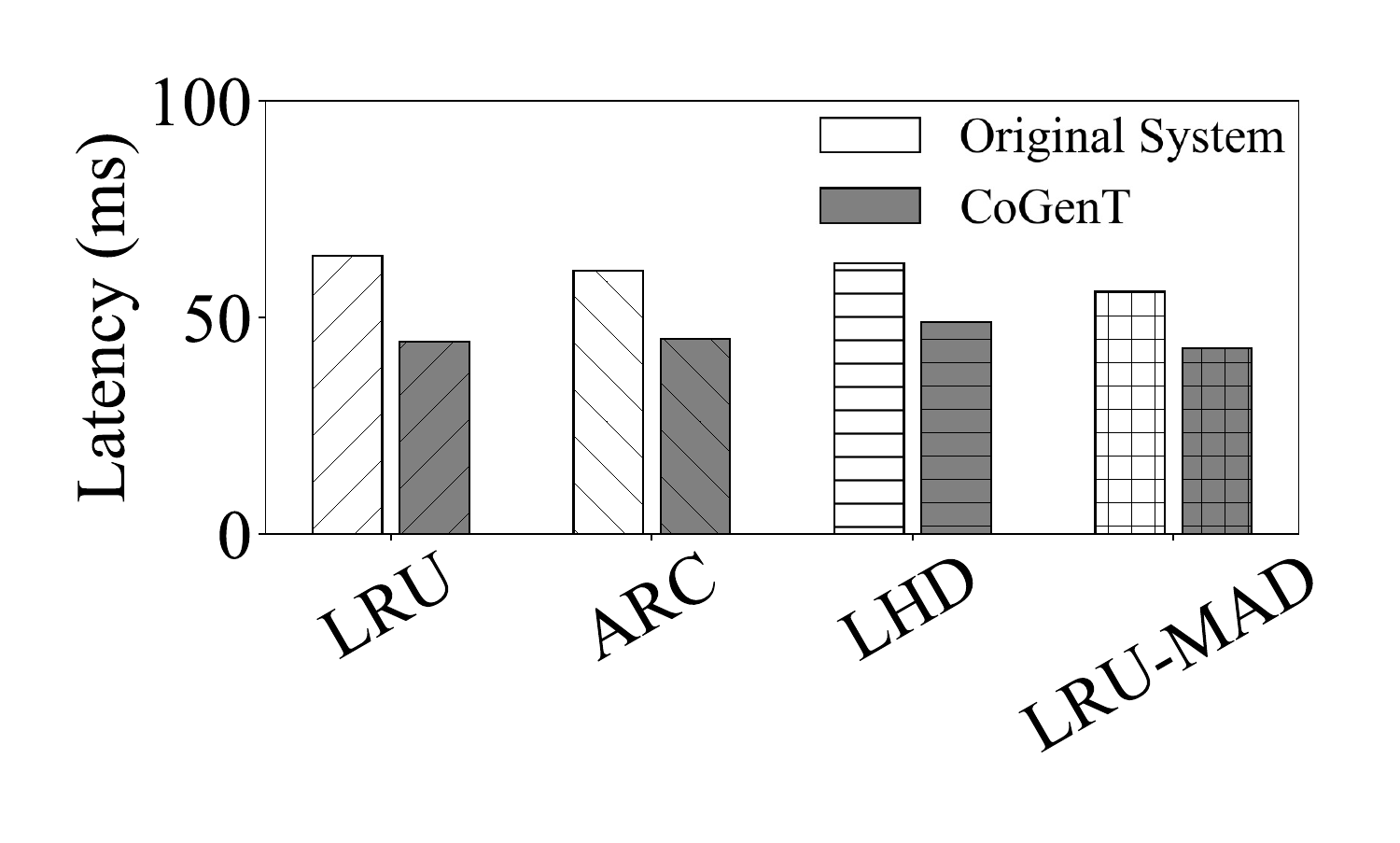}
	    \label{fig:latency2}
	}
    \subfigure[Back-to-COS bandwidth]{
	    \includegraphics[width=0.46\linewidth]{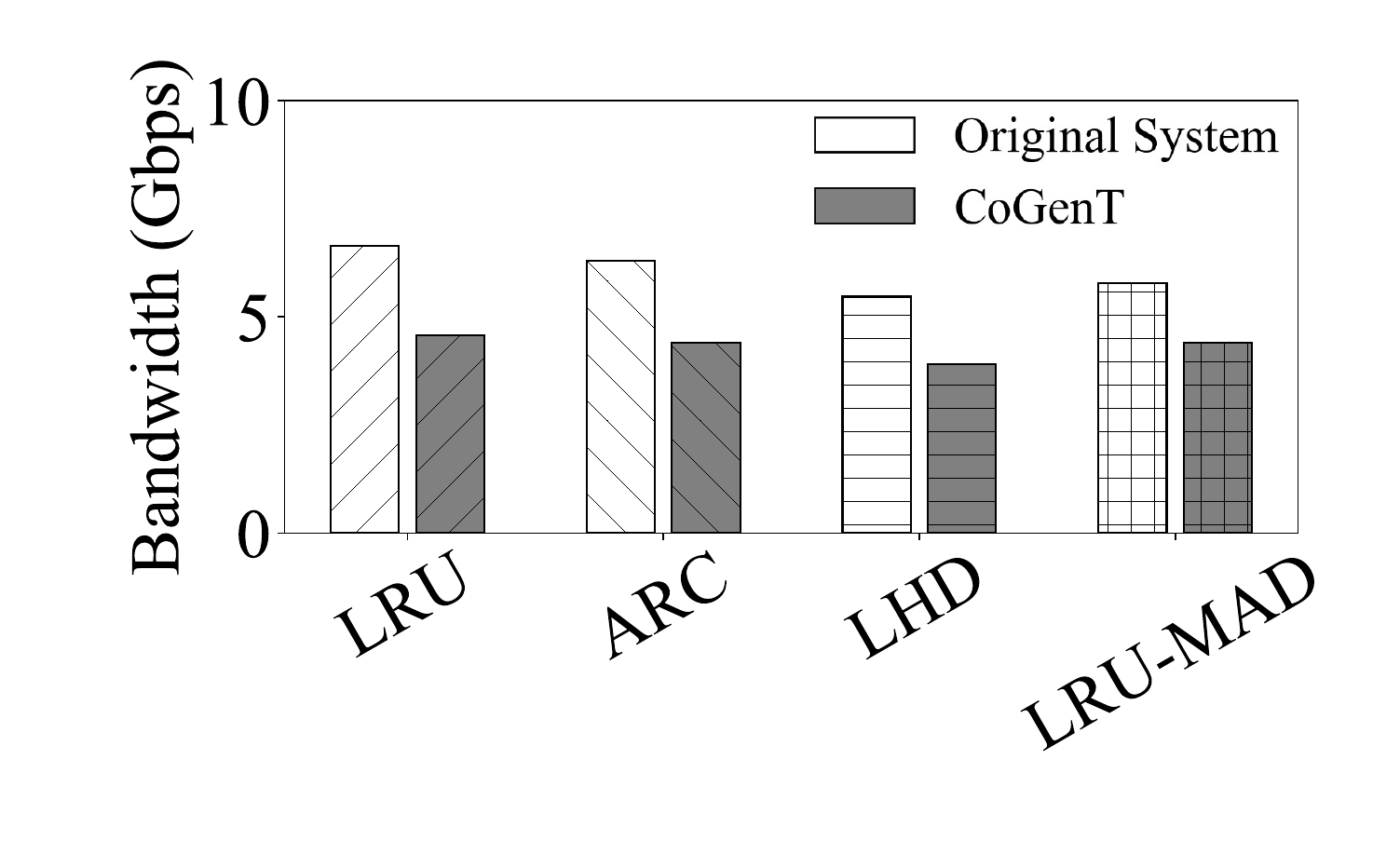}
	    \label{fig:bandwidth2}
    }
    \caption{Average access latency and back-to-COS bandwidth yielded by different cache algorithms on the original system and \fname.}
    \label{fig:org_copm}
\end{figure}

\begin{figure}[t]
    \centering
    \subfigure[Case A]{
	    \includegraphics[width=0.46\linewidth]{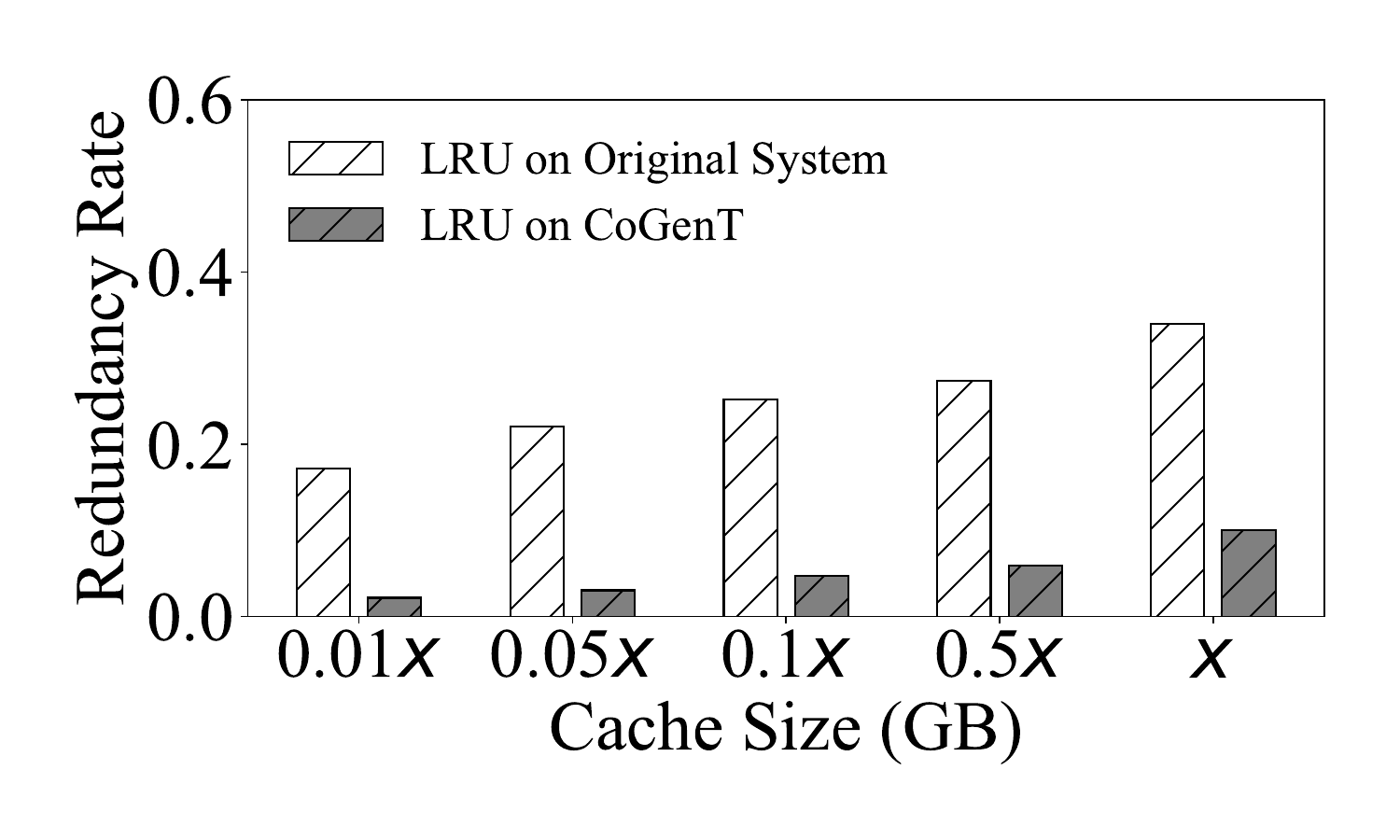}
	}
    \subfigure[Case B]{
	    \includegraphics[width=0.46\linewidth]{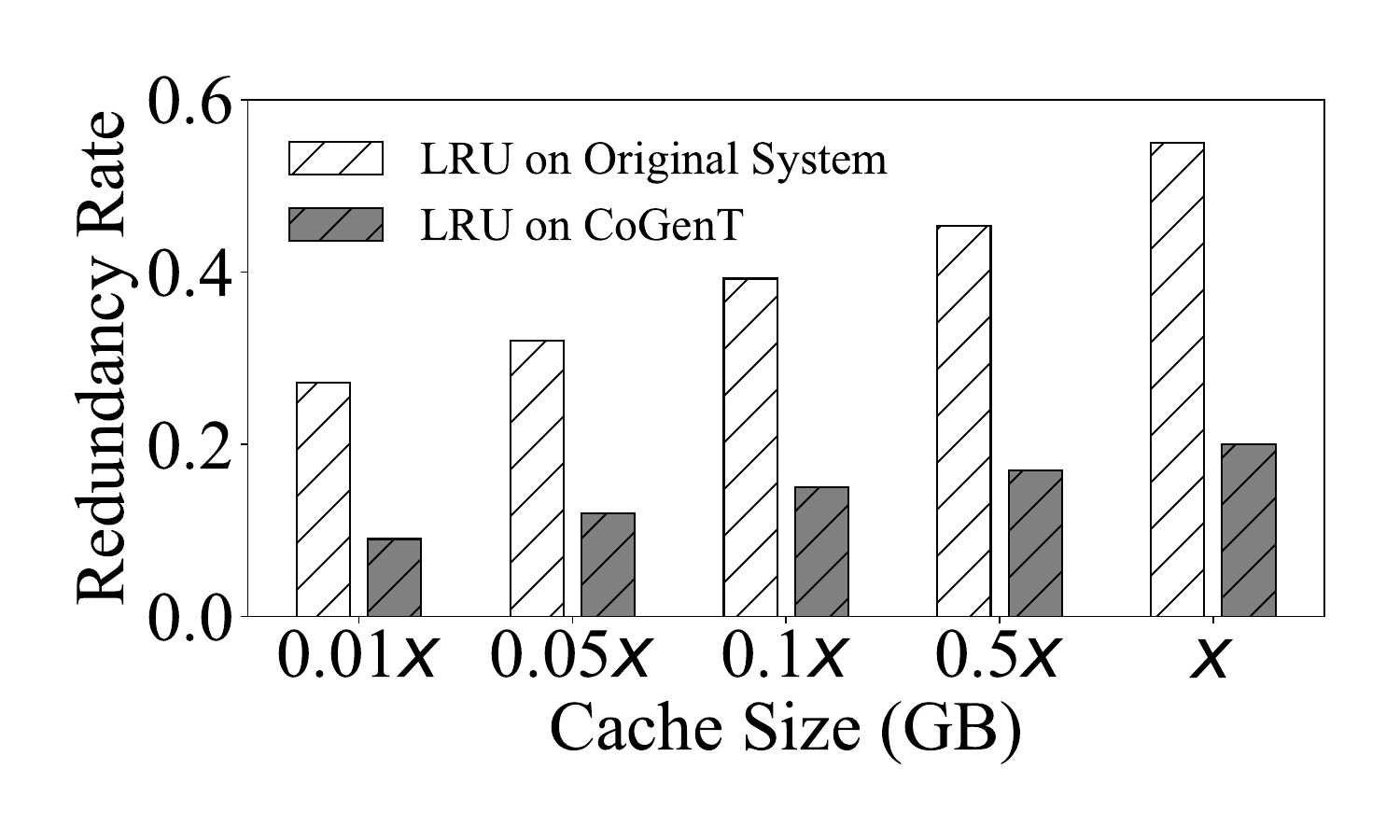}
    }
    \caption{Comparison of content redundancy rates yielded on the simulator. $x=512$GB in Case A and $x=2048$GB in Case B.}
    \label{fig:rr_ab}
\end{figure}

\noindent\textbf{For enhancement effect}, we compare~\fname~with the original system using the same caching algorithm. As shown in Figure~\ref{fig:org_copm}, the latency and back-to-COS bandwidth provided by different algorithms both decrease on~\fname. This suggests that any caching algorithm has the potential to enhance performance on~\fname.

\noindent\textbf{For content redundancy}, we compare our scheme (\ie, LRU running on~\fname) to the baseline solution (\ie, LRU running on the original system) on two open-source datasets~\cite{zhou2018demystifying,liu2023slap}. Note that we do not compare the data deduplication schemes because they do not prioritize access latency. As shown in Figure~\ref{fig:rr_ab}, our scheme achieves an average reduction of 20.0\% and 25.2\% for Case A and Case B, respectively. This reduction involving content redundancy can reduce the number of "invalid writes" of SSDs. In this experiment, the rates of fetching data in the two-pronged control strategy are 5.2\% and 14.6\% for Case A and Case B, respectively. Due to a lack of computing resources, 3.8\% and 6.1\% of pseudo-missing requests are treated as traditional missing requests, respectively.

\section{Related Work}

Improving the quality of experience (QoE) in caching is a well-known topic. Research on traditional cache algorithms, such as replacement and admission algorithms, primarily focuses on data locality. CACHEUS~\cite{rodriguez2021learning} and LHD~\cite{beckmann2018lhd} optimize algorithms using local access rules, while LRB~\cite{song2020learning} and RLB~\cite{yan2020rl} improve hit rates using machine learning models. LRU-MAD~\cite{atre2020caching} is the first to address the delay-sensitive problem. Another approach is to push computation closer to the data in disaggregated storage~\cite{zhang2022compucache}.

\section{Conclusion}
We introduce pseudo-missing scenario, lighten the idea of generative hit, and propose a content-oriented generative-hit framework (\fname) for CDNs.~\fname~achieves this by generating data at the cache layer, using the idle computational resources of the edge nodes. To avoid overloading CPU utilization caused by frequent access for pseudo-missing data, we employ a decision tree model to decide whether to fetch data. This solution provides a new perspective on improving caching QoE via data generation. We believe that the increasing computing resources in edge nodes and the development of large language models will result in greater performance improvements for CDNs using our solution.

%
%
%
%

\end{document}